\documentclass[12pt,letterpaper]{article}
\usepackage[a4paper, total={7in, 10in}]{geometry}

\usepackage{graphicx}
\usepackage{helvet}
\usepackage{authblk}
\usepackage{hyperref}
\usepackage{amsmath} 
\usepackage{amssymb} 
\usepackage{orcidlink} 
\usepackage[super,comma,sort&compress]  
   {natbib}
\usepackage[right]{lineno} \linenumbers

\makeatletter
\renewcommand{\maketitle}{\bgroup\setlength{\parindent}{0pt}
\begin{flushleft}
  \textbf{\@title}
  
  \@author
\end{flushleft}\egroup}
\makeatother

\title{An AI-driven Assessment of Bone Density as a Biomarker Leading to the Aging Law}
\date{}

\author[1, *, $\dagger$]{{Linmi} {Tao}}

\author[1, $\dagger$]{{Donglai} {Tao}}

\author[1]{{Ruiyang} {Liu}}

\author[1]{{Yu} {Cheng}}

\author[1]{{Yuezhi} {Zhou}}

\author[2, **]{{Li} {Huo}}

\author[3, ***]{{Zuoxiang} {He}}

\author[3]{{Ti} {Jiang}}

\author[1]{{Jingmao} {Cui}}

\author[1]{{Yuanbiao} {Wang}}

\author[2]{{Guilan} {Hu}}

\author[4]{{Xiangsong} {Zhang}} 

\author[5]{{Yongwei} {Pan}}

\author[5]{{Ye} {Yuan}}

\author[5]{{Yun} {Liu}}

\affil[1]{{Department of Computer Science and Technology}, {Tsinghua University}, {{Beijing}, {100084}, {China}}}

\affil[2]{{Nuclear Medicine Department}, {Peking Union Medical College Hospital}, {{Beijing}, {100006}, {China}}}

\affil[3,5]{{School of Clinical Medicine}, {Beijing Tsinghua Changgung Hospital}, {{Beijing}, {102218}, {China}}}

\affil[4]{{Department of Nuclear Medicine}, {First Affiliated Hospital of Sun Yat-sen University}, {{Guangzhou}, {510080},  {China}}}

\affil[$\dagger$]{Co-first authors.}

\affil[*]{Correspondence: linmi@tsinghua.edu.cn}
\affil[**]{Correspondence: huoli@pumch.cn}
\affil[***]{Correspondence: zuoxianghe@tsinghua.edu.cn}

\begin{document}

\maketitle

\section*{SUMMARY}

As global population aging intensifies, there is growing interest in the study of biological age. Bones have long been used to evaluate biological age, and the decline in bone density with age is a well-recognized phenomenon in adults. However, the pattern of this decline remains controversial, making it difficult to serve as a reliable indicator of the aging process. Here we present a novel AI-driven statistical method to assess the bone density, and a discovery that the bone mass distribution in trabecular bone of vertebrae follows a non-Gaussian, unimodal, and skewed distribution in CT images. The statistical mode of the distribution is defined as the measure of bone mass, which is a groundbreaking assessment of bone density, named Trabecular Bone Density (TBD). The dataset of CT images are collected from 1,719 patients who underwent PET/CT scans in three hospitals, in which a subset of the dataset is used for AI model training and generalization. Based upon the cases, we demonstrate that the pattern of bone density declining with aging exhibits a consistent trend of exponential decline across sexes and age groups using TBD assessment. The developed AI-driven statistical method blazes a trail in the field of AI for reliable quantitative computation and AI for medicine. The findings suggest that human aging is a gradual process, with the rate of decline slowing progressively over time, which will provide a valuable basis for scientific prediction of life expectancy.

\section*{KEYWORDS}

Bone mineral density, Aging law, Biological indicator, Deep learning

\section*{INTRODUCTION}

``Aging" has been recognized since ancient times. For a long time, we have been seeking for a quantifiable and objective indicator that better reflects the aging process of the human body than chronological age (CA), namely, biological age (BA) \cite{baker1988biomarkers}. Recently, telomeres and DNA have been incorporated \cite{zhang2014select, jylhava2023biological}, resulting in BA models \cite{blackburn2006telomeres} with relatively greater explanatory power. However, \citet{allaire2023genetic} pointed out that differences in leukocyte telomere length (LTL) can be independently influenced by genetic and phenotypic factors, and that both excessively long and excessively short LTL are significantly associated with a reduced average lifespan. 

This suggests that the human body, as a complex system, does not exhibit aging solely through molecular biomarkers but rather through multi-level expressions at the genetic, cellular, and organ levels. Traditional macroscopic indicators are more direct for assessing BA and serve as necessary, easily measurable markers, which was assessed comprehensively through tests of various organ functions\cite{borkan1980assessment}, such as cardiovascular\cite{karavidas2010aging}, lung\cite{skloot2017effects}, liver\cite{gagliano2007mechanisms}, kidney\cite{denic2016structural}, brain and cognitive functions\cite{ran2022brain,ravndal2025sex}, etc.. These indicators themselves are often influenced by factors other than aging \cite{mcclearn1997biogerontologic}. For instance,  \citet{sieren2022menstrual} report that women’s FEV1\%, FVC, and DLCO levels significantly decline during menstruation. \citet{ravndal2025sex} report that brain aging biomarkers, e.g. brain cortical thickness, area and hippocampus volume, show age-dependent sex differences varying among brain regions. The results suggest that these biomarkers cannot fully explain brain aging process. As comprehensive quantifiers of human aging, these indicators fail to establish a clear relationship between chronological age (CA) and indicators \cite{klemera2006new}, making it difficult to understand the laws of human aging. Therefore, it is a constant grand challenge of seeking for a biomarker that changes steadily and significantly over long periods to serve as reliable indicators of human aging.

Bones, as a major organ in the human body, have a relatively stable morphology while undergoing significant changes during the processes of growth and aging. Bone age, determined by the length of the epiphysis, is typically assessed in individuals under 18 years old, with a clear relationship to CA \cite{mughal2014bone}. Bone age can indicate the growth process of adolescents and provide effective guidance for intervening in abnormal growth or development \cite{satoh2015bone}. 

Although epiphyseal length undergoes minimal change after adolescence, bone density shows a significant decline during the aging process in adults \cite{ebbesen1999age}, influenced by factors such as age, sex, and others. However, the specific patterns of bone aging remain debated. For instance, \citet{russo2003aging} reports a linear decline in trabecular bone density in the tibia for both men and women between the ages of 20 and 80, while cortical bone density shows a marked acceleration in its decline after age 40. Similarly, \citet{hung2015age} observed that bone density and the number of trabeculae in the radius and tibia decrease linearly in women in the same age range. In contrast, \citet{macdonald2011age} suggests that changes in the bone density of the radius and tibia differ before and after menopause, while the number of trabecular remains roughly constant after menopause.

These investigations have shown a consensus that bone density declines steadily while the discrepancy among these studies primarily stems from differences in the measuring spots, methods, and precision of bone density measurement, since the calculation of quantitative indicators from the results of clinical examinations is always a challenge in medicine. To address these issues, We proposed a two-stage artificial intelligence (AI)  driven method to precisely segment the trabecular bone in CT images of the lumbar spine, and to establish a distribution model of the trabecular bone voxels for calculating a representative value of bone density (trabecular bone density, TBD) from conventional CT scans. Statistical analysis revealed a relationship between TBD values and CA, which is modeled mathematically. This model indicates that TBD serves as a stable indicator of the human aging process and holds promise for measuring the degree of aging.

\section*{RESULTS}

\subsection*{A novel method for the accurate measurement}

Dual-energy X-ray absorptiometry (DXA) is daily used to measure the bone density of the entire bone\cite{doroudinia2015bone} in hospitals. Quantitative computed tomography (QCT) is mostly applied in researching bone density measurement, where region of interest (RoI) is selected either manually or automatically to represent the bone density\cite{adams2009quantitative} with the average value within the RoI. While end-to-end, AI-based methods\cite{moro2025development, nguyen2025artificial} have been developed to improve the efficiency of DXA or QCT-based bone density measurement, these methods are built on a hidden assumption: the bone mass has a homogeneous distribution in the bone. This assumption is not verified and thus remains questionable.

Based on previous studies \cite{giambini2013longitudinal, oppenheimer2018trabecular}, current diagnostic guidelines, and clinical experience, we chose the trabecular bone of vertebrae L1 and L2 as the representation bone of human body, and collected CT images of L1 and L2 of the patients undergoing whole-body PET/CT examinations from Peking Union Medical College Hospital and the First Affiliated Hospital of Sun Yat-sen University, with a total number of 750 and 891 cases respectively. By excluding slices close to the edge of a vertebrae or containing an intervertebral disc, we collected 5-10 images for each case, according to the scanning slice thickness.

To compare with individuals with bone-related diseases, we collected CT images of 78 patients, who had fractures or other bone diseases, of which 71 cases were diagnosed with reduced bone mass or osteoporosis by DXA, from Beijing Tsinghua Changgung Hospital (BH). We used the same method as previously to select suitable slices for subsequent analysis.

All of the above data were obtained from the database of routine outpatient visits at the hospitals.

A deep learning model is used and specifically designed for the precise segmentation of trabecular bone (Figure 1). This method acquires the whole trabecular bone of L1 and L2, while eliminating the arbitrariness introduced in previous practices when defining RoI, excludes interference from cortical bone and surrounding tissues. 

We use statistical methods, instead of AI-based methods, to analyze the segmented CT pixels. The testing of multiple distribution families reveals that the bone mass in the trabecular bones presents a typical skewed distribution (Figure 2a-2d), Calculations show that the skewness of the distributions of most individuals exceeds 0.99 (Figure 2e), representing a typical skewed distribution. This results in the conclusion that the CT values of the trabecular bone belong to a family of distributions with triple parameters or more, which does not follow the Gaussian distribution used in the assumption of traditional studies. In addition, trabecular bone and cortical bone follow different distribution patterns (Figure 7). Therefore, both the whole-bone measurement in DXA and the RoI selection process in QCT introduce systematic errors. These results demonstrated that the method of combining AI with a mathematical model is emerging from AI for medicine due to its explainability and reliability. 

\subsection*{Trabecular Bone Density}
Based on the precise segmentation results of the trabecular bone, we performed a statistical analysis of the CT values (in Hounsfield Unit, HU) of the trabecular bone pixel by pixel for L1 and L2 individually. After testing multiple distribution families, we select the Johnson-SU distribution, which provides the best fit. Thus, we conclude that the CT values of the trabecular bone of L1 and L2 follow a Johnson-SU distribution pattern.

Mathematically, the representative value of a Johnson-SU distribution is the mode of the distribution, which is defined as the representative value of the TBD of an individual. Figure 2f exemplifies the difference between TBDs and commonly used mean values. 

Research has shown that under the assumption that the composition ratio of non-hydroxyapatite components in the trabecular bone remains approximately constant at a single site, the CT value of the trabecular bone has a linear relationship with the hydroxyapatite content\cite{sudhyadhom2020molecular}. In practice, QCT uses multiple calibration phantoms with different equivalent hydroxyapatite contents to establish the quantitative relationship between CT values and true bone density, demonstrating that with appropriate sampling and analysis, CT values and volumetric bone density obtained by QCT are equivalent. Thus, CT values alone can represent volumetric bone density.

From the principle of bone density measurement, the TBD obtained in this study serves as a measurement of the equivalent hydroxyapatite content in the trabecular bone, which can be used daily to measure bone density in hospitals.

\subsection*{The TBD based aging law in adults}

We used TBD as a biomarker to discover the law of aging in adults since it is calculated with high precision and reliability. We selected 1,018 cases from the collected 1,641 cases for the analysis of aging law. Cases with bone disease or photographic artifacts are excluded, as well as cases with ages below 39 or above 80 due to their limited numbers. We have tested various models of the decaying with aging, and selected the exponential decaying model with the consideration of the microscopic mechanism of bone density loss described by \citet{seeman2019antiresorptive} and the actual data distribution. Using this model, we performed exponential regression analysis by gender and age group (middle-aged group: $39 \leq \text{age} < 60$; elderly group: $\text{age} \geq 60$). Detailed results are shown in Figure 3c and 3d and Table 1. Thus, Equation \ref{eq:aging_law}, where $39 \leq x \leq 80$ is the age and $y$ is the TBD, accurately describe the aging law of the bone density for their respective populations (Figure 3a and 3b).

\begin{equation}
\begin{aligned}
    y_M & = \exp{(5.9987 - 0.0193x)}\\
    y_F & = \exp{(6.5307 - 0.0299x)}
\end{aligned}
\label{eq:aging_law}
\end{equation}

\subsection*{Survivor bias revealed by the aging law}

In both genders, the fitted equations for the elderly group showed lower initial values and slower rates of decline compared to the middle-aged group. This indicates that the measured elderly population had overall higher bone density values compared to those extrapolated from the middle-aged group, which is attributed to systemic errors in data collection (Figure 3e and 3f).  In Figure 3c and 3d, differences are observed in the overall fitting results compared to the middle-aged group, which were primarily caused by a significant elevation in the elderly group’s data. This difference increased with age. We attribute this elevation primarily to survivor bias: individuals with relatively low bone mass may have difficulty participating in routine health check-ups due to mobility issues and are therefore underrepresented in our sample.

To validate this hypothesis, we selected a group of individuals with bone-related diseases from a third-party hospital, specifically choosing those diagnosed with osteoporosis or osteopenia through hip DXA scans. Due to the small sample size, linear fitting is performed after extracting the TBDs, separated by gender (Figure 3e and 3f). For ease of comparison, we linearized the fitting curves of the middle-aged and elderly groups within the 60-90 age range, as shown by the dashed lines in the figures.

It can be clearly observed that the fitting results for the osteoporosis and osteopenia group are nearly parallel to the elderly group’s line, and simultaneously fall below the middle-aged and elderly groups. This fact supports the hypothesis of survivor bias.

\section*{DISCUSSION}

\subsection*{Insights given by the aging law}

This study reveals an exponential decline in bone density with age, while earlier studies typically used bilinear fitting methods\cite{doroudinia2015bone,alvarenga2017age,alvarenga2022age}. From a statistical perspective, the error rates of these methods are essentially comparable to ours: both are able to describe the age-related decline in bone density. Bilinear models are often used to explain the pattern of changes in bone density before and after menopause in women. However, in our model, it is observed that menopausal women exhibit a greater absolute decline in bone density compared to men and postmenopausal women, which justifies the use of a bilinear split point around the ages 50-55. Yet, from a statistical standpoint, there is no consensus on a definitive segmentation point for bilinear models, making this division more of a phenomenological interpretation.

In terms of the decline in relative bone density, both sexes follow a consistent trend (Figure 3c and 3d). Thus, although the change in bone density is a complex, multilayered process influenced by many factors, it can still be interpreted on a macroscopic level as a time-dependent, decay-like process.

This pattern suggests that human aging is a slow and continuous process, rather than one marked by abrupt changes at specific time points. Based on this understanding, the observed trend can be extrapolated to older populations (Figure 4). It can be seen that the absolute rate of bone density decline slows significantly in the elderly. Therefore, it is reasonable to believe that with the introduction of more medical interventions in the future, our expected lifespan may continue to increase.

\subsection*{The principle of bone density decline with aging}

At the microscopic level, changes in bone density are driven by the activity of a large number of Bone Multicellular Units (BMUs) composed of osteoclasts or osteoblasts\cite{seeman2019antiresorptive}, 
generated by bone marrow-derived mesenchymal stem cells (BMSCs). Due to their multipotent differentiation potential into osteoblasts, adipocytes, and chondrocytes, BMSCs play a pivotal role in maintaining skeletal homeostasis\cite{baker2015characterization}. However, with aging, BMSCs undergo cellular senescence, leading to impaired proliferation and skewed differentiation—specifically, a decline in osteoblasto genesis and an increase in adipogenesis\cite{pignolo2021bone}. This shift contributes to the reduced activity of BMUs in bone formation, further leading to bone mineral decrease and marrow adiposity in senile osteoporosis\cite{qadir2020senile}. 

Theoretically, when a large number of BMUs operate independently and the probability of activity for each BMU is very low, the total number of BMU activities follows a Poisson distribution, according to the Poisson limit theorem. The total bone mass is the cumulative result of these cellular activities, which mathematically manifests as exponential decay due to the average activity duration of BMUs. Therefore, we employed exponential decay model based on Poisson regression to address this principle of bone activity with aging, thereby establishing a bridge between microscopic activities and the macroscopic phenomenon of bone density reduction over time.

\subsection*{A Reliable method in AI for quantitative computation} 

 The development of AI has provided tools for analyzing large datasets and modeling examination results. Such approaches have already been applied to the brain\cite{cole2017predicting}, skeleton\cite{spampinato2017deep}, heart\cite{goallec2021dissecting}, liver and pancreas\cite{le2022using}, etc., making considerable progress. However, these end-to-end methods use neural networks to model the relationship between examination results and indicators as a complex function to estimate the values of indicators. This approach inevitably faces interference from input noise, such as the unrelated part in a CT image, which makes the estimated value unreliable for its limited interpretability. Therefore, obtaining quantitative indicators from clinical examination results remains a central challenge in AI for quantitative computation. 

We proposed a two-stage method: an AI-based segmentation model to segment a CT image to two parts, area for analysis and unrelated area to be eliminated, and a statistical method to calculate the value of indicator, which allows us to establish reliable models that connect macroscopic observations to the microscopic aging process, with its reliability guaranteed by mathematics. Consequently, our approach is inherently reliable and provides a feasible pathway to address the reliability challenge faced by AI for medicine.

\subsection*{Conclusion}

By introducing a precise AI-driven bone density measurement method, we have discovered a new assessment of bone density, TBD, and the pattern of bone density changes with age. We observed that bone density goes a steady and gradual exponential decline with aging, and the rate of decline is higher in females than in males.

The measurement method is based on CT imaging, which is easy to perform and accurate. By comparing the obtained results with the aging law, an individual’s biological age can be estimated. Thus, TBD serves as a stable indicator of human aging and can provide a valuable reference for medical diagnosis.

\section*{METHODS}

We collected CT images from a total of 1,641 patients who underwent full-body PET/CT scans from Peking Union Medical College Hospital (PH) and the First Affiliated Hospital of Sun Yat-sen University (FH) in Guangzhou. The details are as follows: (1) 750 cases from PH, with a Siemens Biograph 64 scanner, tube voltage of 120kV, tube current of 128mAs, slice thickness of 3mm, and resolution of $512 \times 512$ (Figure 5a); (2) 891 cases from FH, with a Philips Gemini GXL 16 scanner, tube voltage of 120kV, tube current of 325mAs, slice thickness of 5mm, resolution of $512 \times 512$ and two different fields of view (FOV) for each slice(Figure 5c, 5d). Additionally, we collected CT images of 78 patients who underwent abdominal PET-CT scan from Beijing Tsinghua Changgung Hospital (BH). The images were obtained using a Philips 64-slice PET/CT, with a tube voltage of 120kV, tube current of 100mAs, slice thickness of 1.25mm, and resolution of $512 \times 512$ (Figure 5b). All of the above data were obtained from the database of routine outpatient visits to hospitals.

We manually annotated the trabecular bone region for 152 cases from PH, with a total of 2,606 slices, for precise segmentation model training (PH dataset). These were randomly grouped into the training, validation and test set, with sizes of 1,442, 549 and 615, respectively. In addition to the regular training procedure, we manually annotated 134 cases from FH, with a total of 858 slices for both FOVs (FH-Small and FH-Big dataset), to test the generalization ability of the model. 

When the segmentation model is trained, all the CT images are automatically segmented by the model for further analysis. The segmentation results from the three hospitals have demonstrated the models are fitted to CT images from vairous hospitals. 

\subsection*{Neural Networks for Trabecular Bone Segmentation}

We have tested various neural network architectures, such as UNet\cite{ronneberger2015u}, Deeplab\cite{chen2017deeplab}, Attention UNet\cite{oktay2018attention}, DenseUNet\cite{cao2020denseunet}, and UNeXt\cite{valanarasu2022unext}, which are developed on the JMedSeg framework. All these models are trained with batch size 16, cross entropy loss and learning rate $10^{-4}$, over the PH dataset for 20 epochs. Detailed results are shown in Table 2. It is shown that all these networks are well suited for the bone segmentation task, but UNet shows the most stability and robustness (Figure 5e, 5f). Thus, we adopted the simplest structure, UNet, to extract the trabecular bone region of the lumbar vertebrae in this paper.

\subsection*{Best-Fitting Distributions for Trabecular CT Values of Individuals}

We used SciPy\cite{2020SciPy-NMeth} as the analysis tool to fit the distribution of the segmented trabecular bone region in an individual. We selected the best-fitting distribution from all the 87 univariate continuous distribution functions supported by SciPy.

We observed that the Johnson-SU distribution, the normal inverse gauss distribution (norminvgauss), and the generalized hyperbolic distribution (genhyperbolic) performed similarly and far outperformed other distributions. Among them, the Johnson-SU and norminvgauss used 4 parameters, while the genhyperbolic used 5 parameters. Figure 6b shows the comparison results of both 4-parameter distributions with the normal distribution (using the Johnson-SU distribution as the baseline). As seen, both distributions have significant advantages over the normal distribution. This indicates that, in general, the CT values of trabecular bone in the vertebrae do not follow a normal distribution. 

Given their similar performance, we finally selected the Johnson-SU distribution as the optimal fit to vertebral CT values. Furthermore, we use the mode value of a Johnson-SU distribution, which the value of the spot where the distribution function reaches its peak, as the representation value the corresponding probability distribution. We define this representative  value as the novel measurement of the bone density, named as Trabecular Bone Density (TBD). 

\subsection*{The Dynamics of TBD Decline}

Assuming that, at the microscopic level, individual BMUs engaged in bone resorption and remodeling can be macroscopically approximated as a large number of mutually independent BMUs collectively working to “maintain a constant bone mass”, and their overall activity can be treated as a sum of these independent processes. As time $t$ progresses, the still-active BMUs become inactivated (i.e., undergo aging) at a certain probability rate $p(t)$ at each moment.

Under this assumption, if the number of active BMUs of an individual at time $t=0$ is $N_0$, $N(t)$ ,namely the number of active BMUs at any time $t > 0$, is given by Equation \ref{eq:diff-of-bmd},
\begin{equation}\label{eq:diff-of-bmd}
    \begin{aligned}
    \frac{d}{dt} N(t) &= -p(t) N(t) \\
    N(0) &= N_0 
    \end{aligned}
\end{equation}
whose solution is trivial,
\begin{equation}\label{eq:bmu-to-time}
    N(t) = N_0 e^{-\int p(t) dt}
\end{equation}

Accordingly, the individual's BMD $D(t)$ is determined by the total number of active BMUs $N(t)$, the amount of bone mass $c$ that each BMU can maintain, and the macroscopic bone volume $V(t)$.
\begin{equation}\label{eq:bmd-to-time}
    D(t) = \frac{c \cdot N(t)}{V(t)}
\end{equation}

For trabecular bone, in the absence of trauma, disease, or other significant events, $V(t)$ remains relatively stable during middle and old age. Therefore, $D(t)$, which is measured through TBD, is mainly determined by $p(t)$ alone,
\begin{equation}\label{eq:tbd-to-time}
    D(t) = D_0 e^{-\int p(t) dt}
\end{equation}.
In this model, the microscopical aging process is described by $p(t)$, and its corresponding measurable effect is the decline of $D(t)$ over time.

Yet it is difficult for us to follow a certain individual across a long period, we can sample a group of cases over a wide lifespan. Specifically, we denote the sample group as $S = \{(t_i, D_i)\}$, where each individual is sampled at age $t_i$ with BMD $D_i$. By expanding Equation~\ref{eq:tbd-to-time} to the first order of $t$, we have
\begin{equation}
    \ln \left(\frac{D_i}{D_{0i}}\right) = -\bar{p} t_i - \int_0^{t_i} \Delta p_i(t) dt
\end{equation}

Here, $\bar{p}$ can be interpreted as the shared aging process resulting from general factors such as gender, ethnicity, social environment, etc., while $\Delta p_i(t)$ represents the individual-specific aging process. Thus, we can apply exponential regression to the sampled group to capture the population-level pattern of how the TBD changes over time.

\newpage

\section*{RESOURCE AVAILABILITY}

\subsection*{Lead contact}

Requests for further information and resources should be directed to and will be fulfilled by the lead contact, Linmi Tao (linmi@tsinghua.edu.cn).

\subsection*{Materials availability}

This study did not generate new materials.

\subsection*{Data and code availability}

\begin{itemize}
    \item The segmentation framework (JMedSeg) and datasets (PH, FH-Small and FH-Big) has been deposited at \url{https://github.com/THU-CVlab/JMedSeg} and is publicly available as of the date of publication. All other code and data reported in this paper will be shared by the lead contact upon request.
    \item Any additional information required to reanalyze the data reported in this paper is available from the lead contact upon request.    
\end{itemize}

\newpage

\section*{MAIN FIGURE TITLES AND LEGENDS}

\noindent\includegraphics[width=0.85\linewidth]{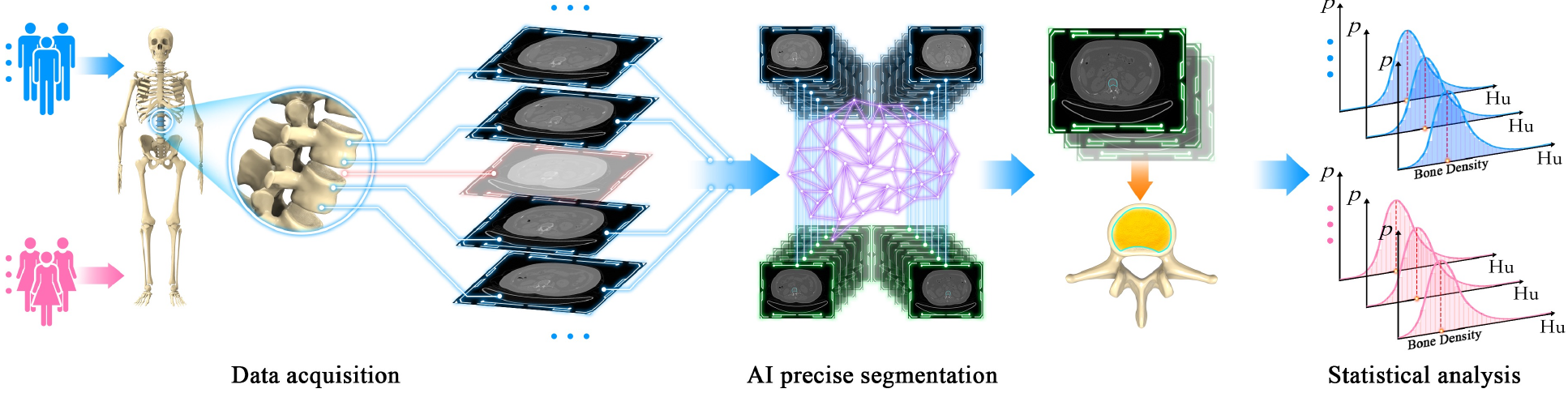}\label{fig:TheProcess}

\subsection*{Figure 1. The AI+statistics method for the mensuration of bone density.}

CT images were obtained by scanning lumbar vertebrae L1 and L2 during the data acquisition stage. The cancellous bone was precisely segmented using the AI-based precise segmentation. An statistical model is fitted to analysis the distribution of segmented bone pixels and the value at the peak of the model is defined as the novel measure of bone density.

\newpage

\noindent\includegraphics[width=1.0\linewidth]{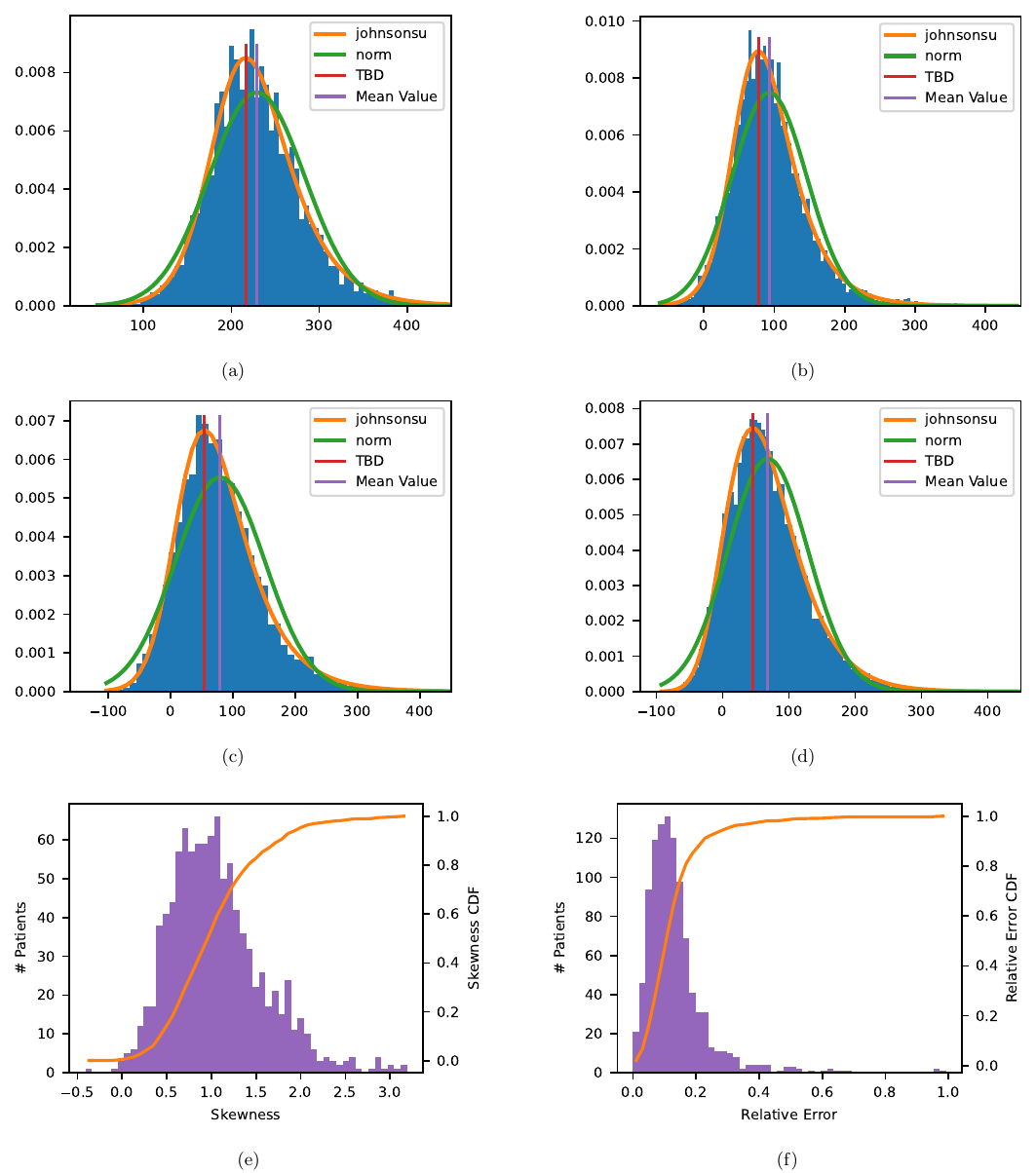}\label{fig:fig2}

\subsection*{Figure 2. The curves fitting to the real distributions.}

Both Johnson-SU (orange curves) and Gaussian (green curves) distributions are fitted to the distributions of L1 and L2 trabecular bone CT values (blue histograms) for individuals across different genders and ages. The corresponding results are shown at the positions of the TBD (red) and mean value (purple). Specifically, (a)\label{fig:fit_M_mid} represents a 45-year-old male, (b)\label{fig:fit_F_mid} a 50-year-old female, (c)\label{fig:fit_M_eld} a 72-year-old male, and (d)\label{fig:fit_F_eld} a 60-year-old female. Additionally, we analyzed (e)\label{fig:total_skewness} the skewness and (f)\label{fig:total_relative_error} the relative error between the mean and the TBD across all samples, finding a median skewness of 0.99 and a median relative error of 13.48\%. These results show that, across all gender and age groups, the  distributions are predominantly skewed, and there is a clear difference between the mean value and the TBD.

\newpage

\noindent\includegraphics[width=1.0\linewidth]{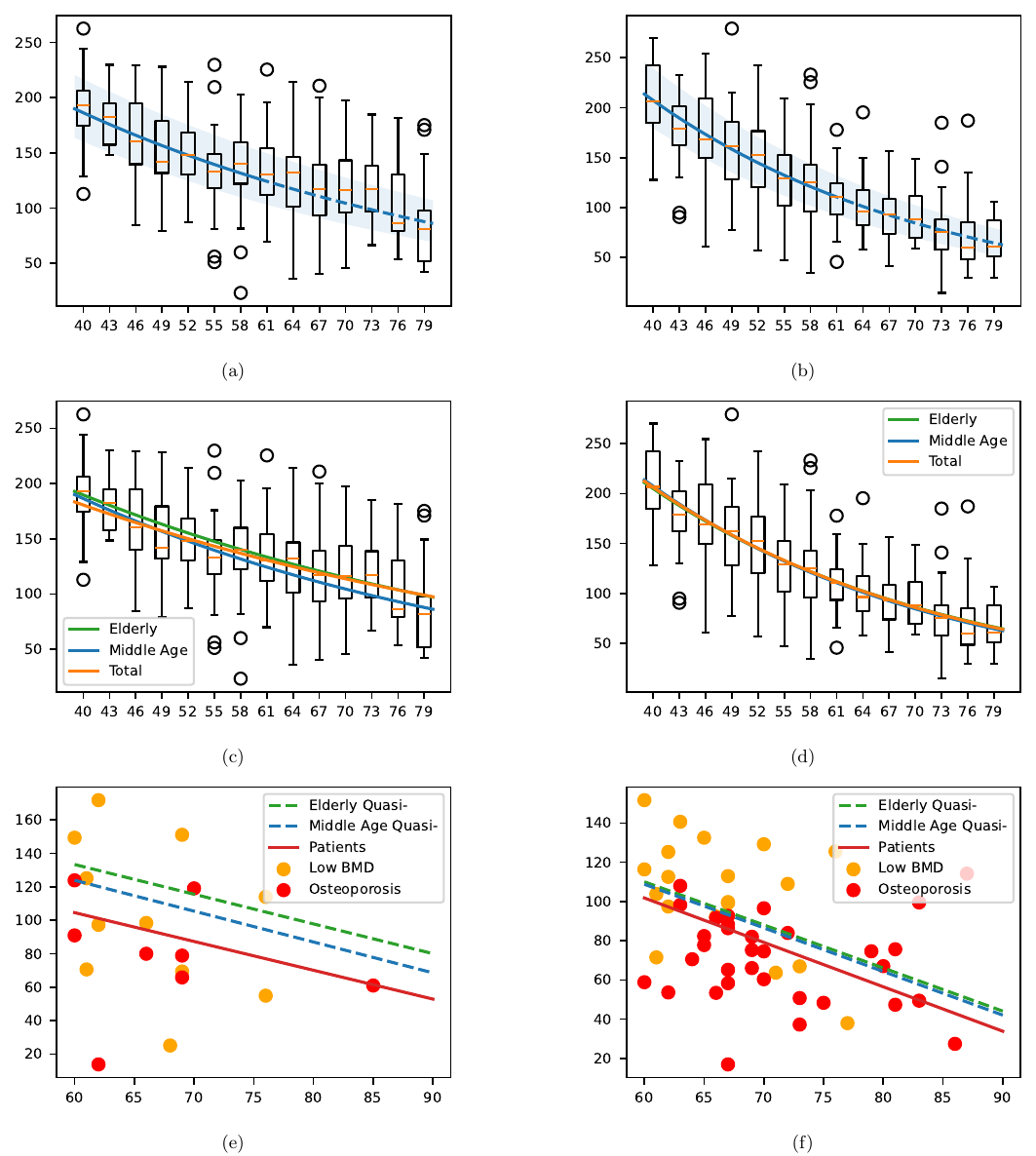}\label{fig:hu_vs_age}

\subsection*{Figure 3. Age-related changes in the trabecular bone of L1 and L2.}

In each figure, the horizontal axis represents age, and the vertical axis represents the TBD (HU). (a)\label{fig:exp_M}, (c)\label{fig:exp_yof_M} and (e)\label{fig:exp_vspa_M} show the fitting results for males, while (b)\label{fig:exp_F}, (d)\label{fig:exp_yof_F} and (f)\label{fig:exp_vspa_F} show the results for females. For the middle-aged group, individuals were grouped into three-year intervals for normalization. Exponential regression was performed on this basis ((a) and (b)) and extended to age 80, with the extended portion represented by dashed lines and the light blue areas indicating confidence intervals. (c) and (d) compare the fitting results for the middle-aged group, the elderly group, and the overall male and female populations, showing a clear shift in the overall results (orange line) toward the elderly group results (green line) at age 80. We attribute this shift to systemic errors in data collection. To investigate further, we compared the fitting results for the middle-aged group, the elderly group, and the elderly population with low bone mass in (e) and (f), which validated this hypothesis. We speculate that this phenomenon is caused by ``survivor bias" in the elderly group. Consequently, we propose that (a) and (b) represent the patterns of change in the TBD of the lumbar spine with age for middle-aged and older males and females, respectively.

\newpage

\noindent\includegraphics[width=0.5\linewidth]{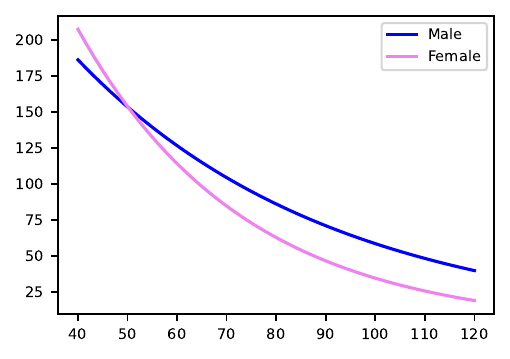}\label{fig:to_120yrs}

\subsection*{Figure 4. Aging law extended to 120 years old people.}

According to Equation \ref{eq:aging_law}, we plotted the expected TBD of people from 40 to 120 years old among both sexes. A significantly slower deline is shown in the elderly. 

\newpage

\noindent\includegraphics[width=1.0\linewidth]{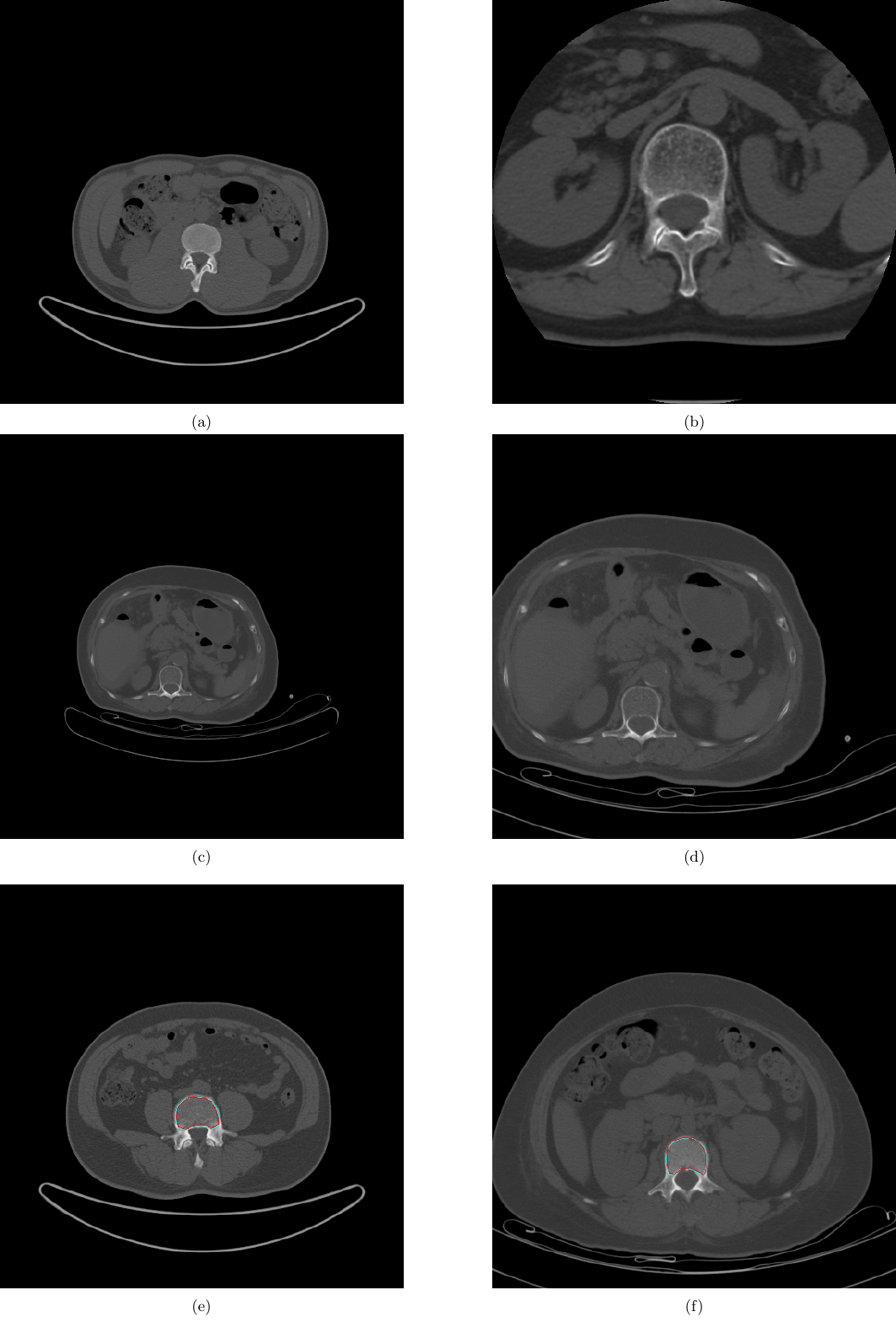}\label{fig:all_compare}

\subsection*{Figure 5. Examples of all three datasets.}

(a-d) Data samples from PH, BH, FH-Small and FH-Big, respectively.
\newline
(e-f) Examples for segmentation results on PH and FH-Big datasets, respectively. Manually annotated POIs (ground truth) are presented as red polygons, and segmentation results given by UNet are presented as cyan polygons.
\newpage

\noindent\includegraphics[width=1.0\linewidth]{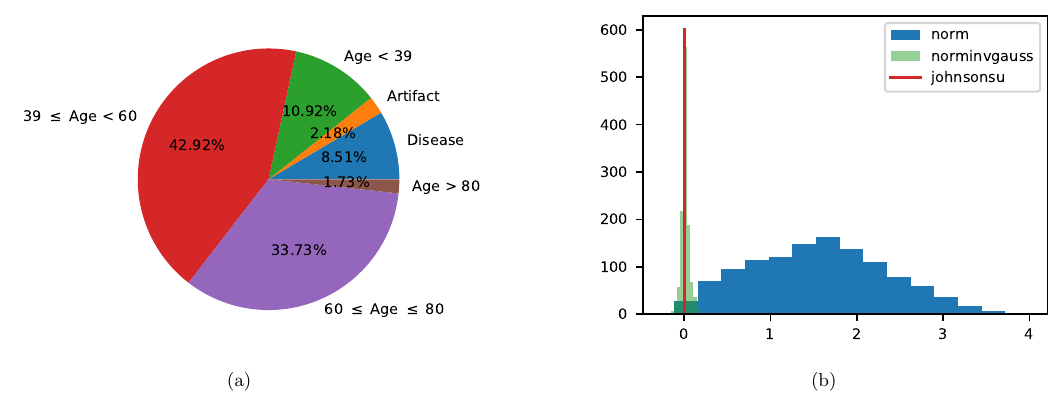}

\subsection*{Figure 6. Distribution analysis of PH and FH datasets.}
(a) Age Distribution of all cases from PH and FH.
\newline
(b) Comparison of MSEs for Johnson-SU, Normal Inverse Gaussian and Normal Distributions cross all 1,018 individuals. For each individual, the MSE of the Johnson-SU distribution is used as the baseline. The ratio of the MSEs of the other distributions to the baseline result is taken, and the natural logarithm of this ratio (on the horizontal axis) is plotted as a histogram.
\newpage

\noindent\includegraphics[width=1.0\linewidth]{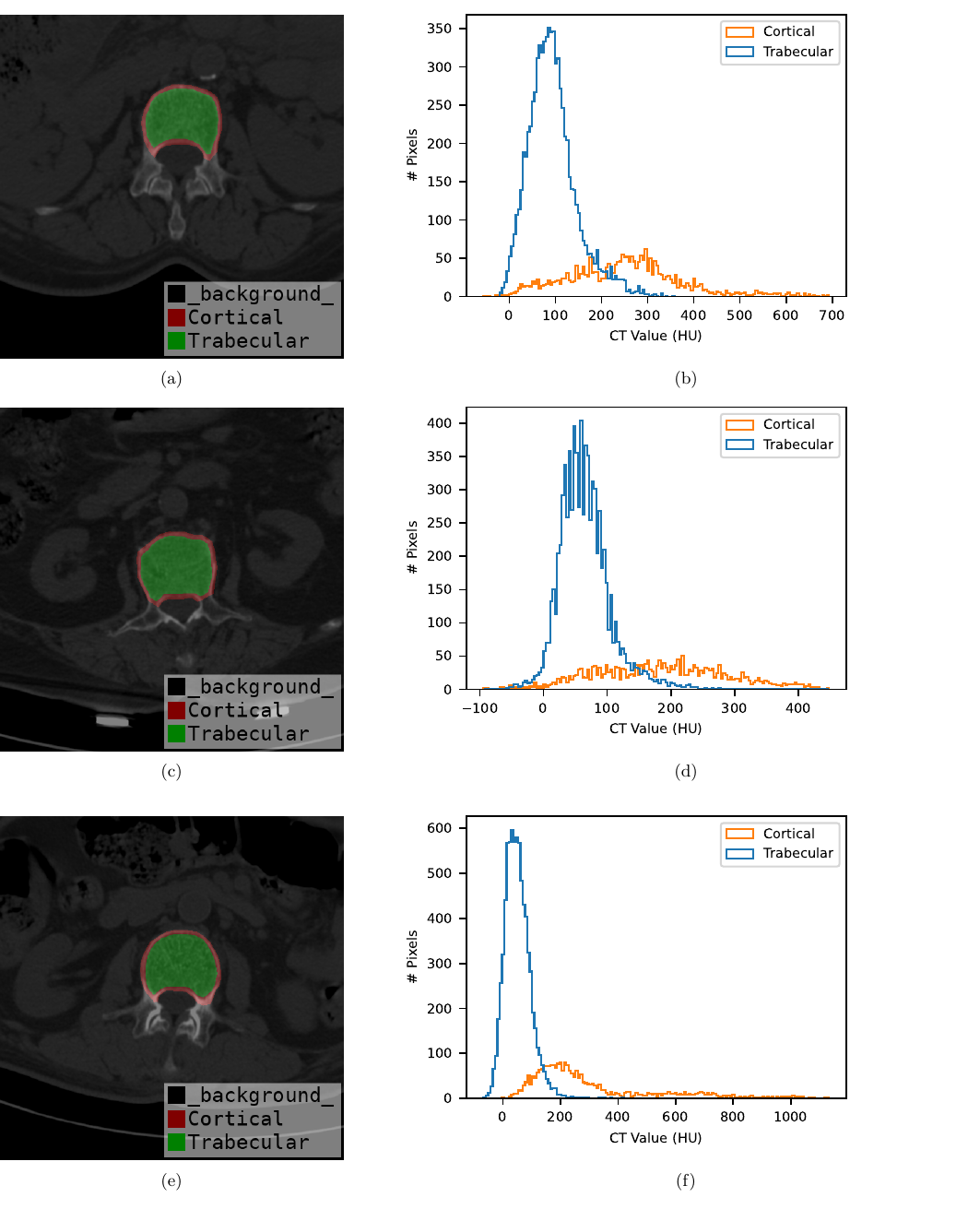}

\subsection*{Figure 7. Distribution of CT values in cortical and trabecular bone regions.}
(a–f) Histograms showing the pixel-wise distribution of CT values for cortical bone (orange) and trabecular bone (blue) from different patients. The x-axis represents CT value (HU), and the y-axis denotes the number of pixels. Across all panels, cortical bone exhibits a higher mean CT value and a broader distribution compared with trabecular bone, reflecting differences in mineral density and microstructural composition. These distributions indicate the necessity of seperating trabecular bone and cortical bone in the analysis of bone density.
\newpage

\section*{MAIN TABLES, INCLUDING TITLES AND LEGENDS}

\subsection*{Table 1. Corresponding fitting parameters.}

\begin{tabular}{|l|l|l|l|l|}
    \hline
    & $y=\exp{(a+bx)}$ & $a$ & $b$ & Pseudo $r^2$\\
    \hline
    Male & Middle-age (39-59 yrs) & $5.9987 \pm 0.043$ & $-0.0193$ & 0.8410\\
         & Elderly (60-80 yrs) & $5.9139 \pm 0.074$ & $-0.0167$ & 0.6120\\
         & Total & $5.8119 \pm 0.0023$ & $-0.0154$ & 0.9515\\
    \hline
    Female & Middle-age (39-59 yrs) & $6.5307 \pm 0.041$ & $-0.0299$ & 0.9892\\
           & Elderly (60-80 yrs)& $6.4781 \pm 0.094$ & $-0.0289$ & 0.8873\\
           & Total & $6.4975 \pm 0.023$ & $-0.0292$ & 1.0000\\
    \hline
\end{tabular}\label{tab:hu_vs_age}

\newpage

\subsection*{Table 2. Test Results of Different Models.}

\begin{tabular}{|l|l|l|l|l|}
    \hline
    Model & Metric & PH & FH-Big & FH-Small \\
    \hline
    UNet\cite{ronneberger2015u} & Dice & 0.9732 &  \textbf{0.9804}* & \textbf{0.9662}* \\
         & IoU & 0.9420 &  \textbf{0.9623} & \textbf{0.9366} \\
    \hline
    Deeplab\cite{chen2017deeplab} & Dice & 0.9711 & 0.9569 & \textbf{0.9620} \\
            & IoU & 0.9453 & 0.9205 & \textbf{0.9293} \\
    \hline
    Attn. UNet\cite{oktay2018attention} & Dice & \textbf{0.9755}* & \textbf{0.9783} & 0.8949 \\
               & IoU & \textbf{0.9533} & \textbf{0.9583} & 0.8261 \\
    \hline
    DenseUNet\cite{cao2020denseunet} & Dice & \textbf{0.9734} & 0.9734 & 0.9453 \\
              & IoU & \textbf{0.9494} & 0.9495 & 0.9014 \\
    \hline
    UNeXT\cite{valanarasu2022unext} & Dice & 0.9650 & 0.9679 & 0.9351 \\
          & IoU & 0.9345 & 0.9397 & 0.8850 \\
    \hline
\end{tabular}\label{tab:model_test_results}

\bigskip

* Top-1 Models.

\newpage

\bibliography{references}

\end{document}